\documentclass[twocolumn, pre, aps, 10pt, superbib, floats,floatfix]{revtex4}

\usepackage{ifthen}
\usepackage{ifpdf}
\usepackage{color}

\ifpdf
\usepackage{graphicx}
\usepackage{epstopdf}
\else
\usepackage{graphicx}
\usepackage{epsfig}
\fi

\usepackage{latexsym}
\usepackage{amsmath}
\usepackage{amssymb}
\usepackage{bm}

\usepackage{hyperref}

\newcommand{\tbox}[1]{\mbox{\tiny #1}}

\newcommand{\pd}[2]{\frac{\partial #1}{\partial #2}}
\newcommand{\ppd}[2]{\frac{\partial^2 #1}{\partial #2^2}}
\newcommand{\fd}[2]{\frac{\mathrm{d} #1}{\mathrm{d} #2}}

\newcommand{\be}[1]{\begin{eqnarray}\ifthenelse{#1=-1}{\nonumber}{\ifthenelse{#1=0}{}{\label{e#1}}}}
\newcommand{\ee}{\end{eqnarray}}

\newcommand{\hide}[1]{}

\begin{document}

\title{Stochastic memory: memory enhancement due to noise}
%\shorttitle{Stochastic memory}

% \date{\today}

\author{Alexander Stotland and Massimiliano Di Ventra}

\affiliation{Department of Physics, University of California--San Diego, La
Jolla, California 92093-0319, USA}

\begin{abstract}

There are certain classes of resistors, capacitors and inductors that, when subject to a periodic input of appropriate
frequency, develop hysteresis loops in their characteristic response. Here, we show that the hysteresis of
such memory elements can also be induced by white noise of appropriate intensity even at very low frequencies of the external driving field. We illustrate this phenomenon
using a physical model of memory resistor realized by $\mathrm{TiO_2}$ thin films sandwiched between metallic electrodes, and discuss under which conditions this effect can be observed
experimentally. We also discuss its implications on existing memory
systems described in the literature and the role of colored noise.

\end{abstract}

\maketitle

%%%%%%%%%%%%%%%%%%%%%%%%%%%%%%%%%%%%%%%%%%%%%%%%%%%%%%%%%%%%%%%%
%%%%%%%%%%%%%%%%%%%%%%%%%%%%%%%%%%%%%%%%%%%%%%%%%%%%%%%%%%%%%%%%

\section{Introduction}
Memory effects are very common in nature and develop whenever the dynamical
properties of a system depend strongly on its history~\cite{pershin:memoryreview}. In certain structures of condensed matter physics these memory features appear most strikingly in the hysteresis behavior of their resistive,
capacitive, and inductive characteristics when subject to time-dependent perturbations. In particular, this hysteresis
is more pronounced for periodic perturbations of appropriate frequencies corresponding to the inverse response time of
some state variables of the system~\cite{pershin:memoryreview}. These memory elements are usually called memristors~\cite{Chua:memristor}, memcapacitors and meminductors~\cite{diventra:memelements}, respectively. Their advantage is that they may retain information without the need of a power source, and they have found
application in diverse areas of science and technology ranging from information processing to biologically-inspired
systems~\cite{driscoll09b,pershin09b,Lehtonen09a,driscoll09a,Gergel09a,pershin10c,pershin10d,jo10a,Driscoll10a}. So far, however, all these studies have neglected the important effect of noise on the memory properties of these elements.

Noise comes in various forms and it can be generally classified as internal or
external to the system \cite{Dutta-Horn-RMP}. While internal noise can
provide much information on the system dynamics, external noise is generally considered a nuisance for practical applications. We then naively expect it to be a detrimental effect on the hysteresis of memory elements.

In this paper we show instead that under specific conditions on the strength of
the noise the latter may induce a well defined hysteresis even at low
frequencies of the driving field. This phenomenon is reminiscent of the
stochastic resonance effect that has found widespread use in research
\cite{Benzi:sr_orig, Gammaitoni:sr, Wellens:sr}.
The optimization phenomenon of hysteretic structures due to the stochastic
resonance has already been reported earlier with regard to bistable
and multistable potentials
\cite{hist_noise_bistable, hist_noise_multistable}, and neural networks
\cite{hist_noise_neural} and has also been formally studied in
\cite{hist_noise_formal}.
In this paper we will make a connection between the ``stochastic memory'' effect
we describe here and the theory of
stochastic resonance. In order to illustrate the effect we predict, we refer to a widely used model of $\mathrm{TiO_2}$ memory resistor in which the motion of oxygen vacancies in the $\mathrm{TiO_2}$ thin film induced by an external
electric field is thought to be responsible for memory \cite{Strukov:memristor}. In this case, the noise could be tuned via, e.g., a temperature variation, thus allowing an easy test of our theoretical predictions. In general,
we anticipate that the phenomenon we discuss may emerge in many experimentally realized
memory systems, therefore providing valuable information on the physical processes responsible for memory.

\section{Stochastic memory elements}

Let us then start by defining a general system with memory. If $u(t)$ and $y(t)$ are any two complementary constitutive circuit variables
(current, charge, voltage, or flux) denoting input and output
of a system, respectively, and $x$ is an $n$-dimensional vector of internal
state variables, a generic memory element is defined by the set of equations~\cite{diventra:memelements}
\begin{eqnarray} \label{eq:gen_def_base}
  y(t)&=& g (x, u, t) u(t) \\
  \dot x &=& f (x, u, t)\,.
\end{eqnarray}
Her, $g$ is a generalized response, $f$ is a continuous $n$-dimensional
vector function, and the dot denotes a time derivative. In particular, the
relation
between current and voltage defines a
memristive system~\cite{Chua:memristor}, the relation between
charge and voltage specifies
a memcapacitive system~\cite{diventra:memelements}, and the flux-current relation gives rise to a
meminductive system~\cite{diventra:memelements}. The distinctive feature of memory elements is the existence of a hysteresis loop
that emerges
when the response $y(t)$ is plotted vs the input $u(t)$ for at least one
cycle (see Fig.~\ref{fig:memristance}). The shape of the hysteresis and its characteristics are determined by
the system itself, initial conditions and the applied input, in particular its shape, frequency
and amplitude~\cite{pershin:memoryreview}. However, quite generally, the hysteresis loop is negligible at very small frequencies (the state
variables are able to follow the external perturbation), and at high frequencies (the state
variables do not have time to adjust to the instantaneous value of the external perturbation). At intermediate frequencies, dictated by the internal
physical processes responsible for memory, a finite hysteresis amplitude develops.

If the state variables are now subject to noise we can extend the above definition as follows~\cite{pershin:memoryreview}~\footnote{Noise introduced directly in the input $u$ would also
produce the phenomenon of stochastic memory. In the present work, however, we consider only the case of
a noiseless source.}
\begin{eqnarray} \label{eq:gen_def}
  \{y(t)\}_\xi &=& \{g (x, u, t)\}_\xi u(t) \label{eq3} \\
  \dot x &=& f (x, u, t) + H(x,u,t) \xi(t) \label{eq4}
\end{eqnarray}
where $\xi(t)$ is an $n$-dimensional vector of noise terms defined by
\begin{eqnarray}
  \langle \xi_i(t) \rangle = 0, \ \
  \langle \xi_i(t) \xi_j(t') \rangle = k_{ij}(t,t'), \ \
  i,j = 1,\ldots, n\,. \ \
\end{eqnarray}
The symbol $ \langle \cdot \rangle$ indicates ensemble average and
$k_{ij} (t, t')$ is the autocorrelation matrix. The symbol $\{\cdot\}_\xi$ then has the meaning of a realization of the stochastic process $\xi(t)$. For white
noise on each independent state variable we can choose
$k_{ij} (t, t') = \Gamma_i \delta_{ij} \delta(t-t')$ with $\Gamma_i$ the
strength of the noise. For colored noise, one would have different
autocorrelation
functions (see below). The term $H(x, u, t)$ is some $n \times n$
matrix function allowing for possible coupling of the noise components.
Clearly, the system is deterministic if $H(x,u,t) \equiv 0$.
Equations~(\ref{eq3}) and~(\ref{eq4})
suggest that at a given frequency of the external input $u(t)$, noise would simply
destroy the hysteresis by introducing random fluctuations in the state variables, and hence in the response function $g$. We give below reasons why this is
not always the case. Instead, under specific conditions, white noise and even more so colored noise assist in the development of the hysteresis loop.
We first provide simple analytical arguments followed by a physical example we have solved numerically.

\section{Results and discussion}
To develop an analytical understanding of this phenomenon let us consider only
memory elements without an explicit
dependence of the response function $g$ and the function $f$ on time. For
simplicity, we assume only one state variable $x$ and a constant coupling of
$H(x,u,t) = 1$ to white Gaussian noise $\xi(t)$ of strength $\Gamma$.
From Eq.~(\ref{eq:gen_def}) we then obtain the equation for the time
derivative of the response function
\begin{eqnarray}\label{eq:dgdtgen}
  \dot g = \pd{g}{x}f(x,u) + \pd{g}{u}\dot u+ \pd{g}{x}\xi(t)
    + \frac{1}{2}\Gamma\ppd{g}{x},
\end{eqnarray}
which has to be interpreted as a stochastic differential equation. The last
term in Eq.~(\ref{eq:dgdtgen}) comes from It\^o calculus \cite{Kloeden:sde}.
Let us consider a sinusoidal field $u(t) = u_0 \sin (\omega_0 t)$
of frequency $\omega_0$ and amplitude $u_0$. We also assume that the internal state variable $x$ is confined between two boundaries
$x_1\leq x \leq x_2$ (consequently, the response $g$ varies between two extreme
values set by these limiting states of the system),
and there is a time scale $t_0$ associated with the change of the internal
state from $x_1$ to $x_2$. Therefore, we expect
that at low frequencies, $\omega_0 \ll 2\pi/t_0$ and for small amplitudes $u_0$, the
state variable follows the slow
change of the input $u(t)$, so that $\pd{g}{x} $ can be approximated, to first order, as
a constant (call it $a$).
Then Eq.~(\ref{eq:dgdtgen}) reads
\begin{eqnarray}\label{eq:responsediff}
  \dot g = a f(x,u) +\pd{g}{u}u_0 \omega_0 \cos\omega_0 t + a\xi(t)
\end{eqnarray}
This equation, along with the above assumptions, reminds us of the phenomenon of
stochastic resonance of a classical particle in a double-well potential
$V_0(z)$ in the large damping limit (see, e.g., Ref. \cite{Wellens:sr} and references
therein). In this case, the
equation of motion for the position $z$ of the particle is
\begin{eqnarray}\label{eq:doblewell}
   \dot z = - \fd{V_0(z)}{z} - \frac{V_1}{c}\cos\omega t + \xi(t)
\end{eqnarray}
where $V_0(z)$ describes the symmetric double-well potential with $2c$ the distance between the
wells minima (equilibrium states of the particle), $V_1 \frac{z}{c}\cos\omega t$ is a
small modulation of the potential, and $\xi(t)$ is white noise. The term $V_1 < \Delta V$
is the amplitude of the periodic modulation of
the potential which is smaller than the potential barrier $\Delta V$ between the two wells.
Hence the periodic driving alone cannot induce transitions from one well to
another. In contrast, there exists an optimal noise strength such that the stochastically activated transitions
between the wells are most likely to occur after one half of a cycle of the
periodic modulation. Consequently, the response is optimally synchronized with
the external driving at some finite noise strength and almost-periodic transitions are observed~\cite{Wellens:sr}.
By comparing Eq.~(\ref{eq:doblewell}) with Eq.~(\ref{eq:responsediff}) we can then associate the function $f(x,u)$ in Eq.~(\ref{eq:responsediff}) with the
force $- \fd{V_0(z)}{z}$ applied to the particle, and the small modulation of the potential with the
term proportional to the periodic field that attempts to drive the system from
state $x_1$ to state $x_2$ and vice versa. However, without the addition of some
noise the driving field is unable, within a period, to induce  transitions
between the response extrema.
This analogy \footnote{Strictly speaking the analogy should
be made with a continuous system. For more details about the stochastic resonance effect in continuous systems see Ref. \cite{McNamara:SR}.} allows us to anticipate
that, like in the case of the stochastic resonance, the hysteresis in the
response $g$ may be produced as a {\it cooperative}
effect between the noise and the
driving at appropriate noise strengths and under the conditions specified above.

%%%%%%%%%%%%%%%%%%%%%%%%%%%%%%%%%%%%%%%%%%%%%%%%%%%%%%%%%%%%%%%%

\begin{figure}[t]
 \centering
 \includegraphics[clip, width=0.45\hsize]{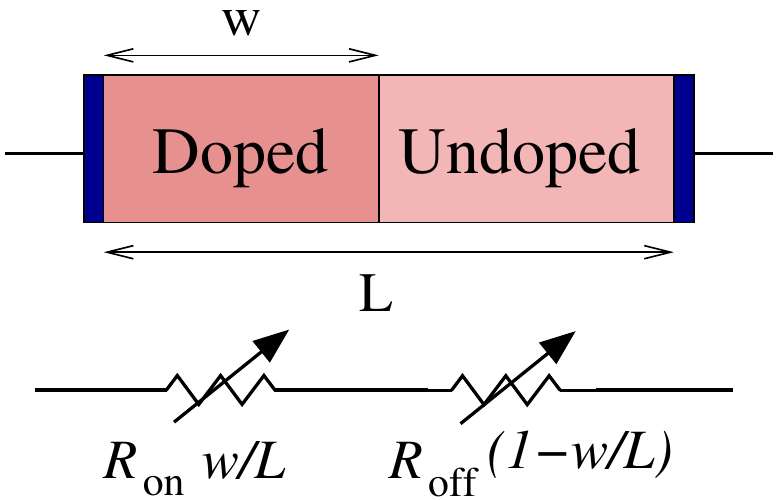}
 \caption{Schematic of a $\mathrm{TiO_2}$ memristor of length $L$ made of a doped and an undoped region, and its simplified equivalent
circuit. The doped region of length $w$ has resistance $\frac{w}{L}R_{\tbox{on}}$ and the
undoped region has resistance $\left(1 - \frac{w}{L}\right)R_{\tbox{off}}$.}
\label{fig:model}
\end{figure}

In order to illustrate the above phenomenon in an experimentally realizable system, we consider the model of a memristor presented in
Ref. \cite{Strukov:memristor} (Fig.~\ref{fig:model}).
It consists of a thin film (on the order of a few nanometers)
containing one layer of insulating $\mathrm{TiO_2}$ and one layer
of oxygen-poor $\mathrm{TiO_{2-x}}$ between two metal contacts. Oxygen
vacancies in the system act as mobile $+2$ charged dopants. These dopants
create a doped $\mathrm{TiO_2}$ region whose resistance is much lower than the
resistance of the undoped region. The location of the boundary between the doped
and undoped regions, and therefore the effective resistance of the film,
depends on the position of the dopants and can be modified by
applying an external electric field. The mobility of the dopants is
${\mu_{\tbox{D}} \sim 10^{-10}\,\mathrm{cm^2 V^{-1} s^{-1}}}$.
We take as total size of the memristor $L= 10\,\mathrm{nm}$ and
denote the length of the doped region as $w$. The current is
$I(t) = {V(t)}/{M(w)}$
with the effective resistance of the system approximated as \cite{Strukov:memristor}
\begin{eqnarray}\label{eq:memristance}
 M(w) &=& \frac{w}{L}R_{\tbox{on}} + \left(1 - \frac{w}{L}\right)R_{\tbox{off}}
\end{eqnarray}
where $R_{\tbox{on}}$ is the resistance of
the memristor if it is completely doped, and $R_{\tbox{off}}$ is its resistance
if it is undoped. In the following we set $R_{\tbox{on}}=1\,\mathrm{k\Omega}$
and $R_{\tbox{off}} =4\,\mathrm{k\Omega}$.
The system is controlled by the applied voltage
$V(t) = V_0 \sin \omega_0 t$.
The memristive effect in this model arises from the time dependence of the
width of the doped region $w$ \cite{Blanc:mreqn} \footnote{For the sake of
illustration, this model omits the diffusion term of the drift-diffusion equation
\cite{Strukov:small}.}
\begin{eqnarray}\label{eq:motion}
 \fd{w}{t} &=& \frac{\mu_{\tbox{D}} R_{\tbox{on}}}{L}
I(t)F\left(\frac{w}{L}\right)  + L \xi(t)
\end{eqnarray}
where the width is limited by $0<w<L$ and $F(x) = 1 - (2x-1)^{2}$ suppresses the speed
of the boundary between the two regions at the edges $w \sim 0$ and
$L$ \cite{Joglekar:memristor}.
The time scale $t_0 = \frac{L^2}{\mu_{\tbox{D}} V_0}$ is the time required for
the
dopants to travel a distance $L$ under a constant voltage $V_0$.

Let us first assume white Gaussian noise of intensity $\Gamma$
\begin{eqnarray}
  \langle \xi(t) \rangle &=& 0,  \ \ \ \ \
  \langle \xi(t) \xi(t') \rangle = \Gamma \delta(t-t'),
\end{eqnarray}
which can be tuned experimentally by varying the temperature
$T$ of the environment. In fact, using the above definitions the diffusion coefficient
is
\begin{eqnarray}\label{eq:diffgamma}
  D = \frac{\Gamma L^2}{2}
\end{eqnarray}
which has to be compared with
\begin{eqnarray}\label{eq:difftemp}
  D = D_0 \exp\left(-\frac{E_\nu}{k_{\tbox{B}} T}\right)
\end{eqnarray}
where $D_0 \approx 10^{-3} \,\mathrm{cm^2/s}$ and $E_\nu \approx 0.5
\mathrm{eV}$ is the activation energy for oxygen vacancy diffusion~\cite{Radecka:diffusionTiO2}.
We note here that the noise in our model was introduced in the simplest possible way such that it shakes the sharp boundary between the doped and undoped regions of the $\mathrm{TiO_2}$ memristor. 
Of course, in the real system the boundary is smeared.
However, it was not our intention to present a detailed physical model of the $\mathrm{TiO_2}$ memristor,
rather to point out an interesting phenomenon due to noise.

\begin{figure}[htp]
 \centering
 \includegraphics[clip, width=0.95\hsize]{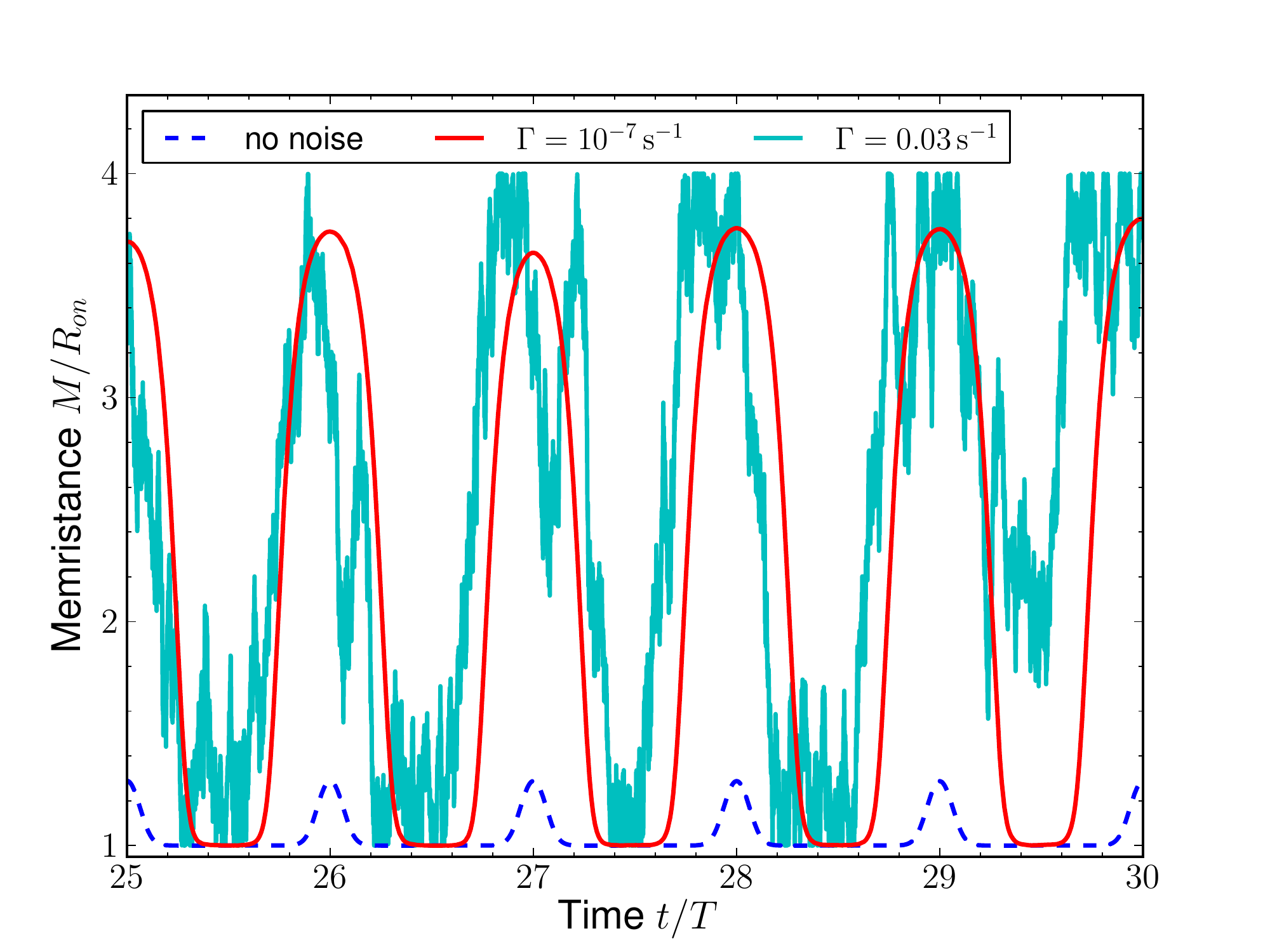}
 \caption{Single realization of the memristance $M(w)$ as a function of time (in units of the time period $T=2\pi/\omega_0$) for different noise strengths. 
The amplitude of the applied voltage is $V_0 = 5.5\,\mathrm{mV}$ and the angular frequency is $\omega = 0.25\,\mathrm{Hz}$. The resistance ratio is
${R_{\tbox{off}}}/{R_{\tbox{on}}} = 4$ and the initial location of the boundary for all simulations is taken as $w_0/L=0.9$.}
\label{fig:memristance}
\end{figure}

\begin{figure}[htp]
 \centering
 \includegraphics[clip, width=0.95\hsize]{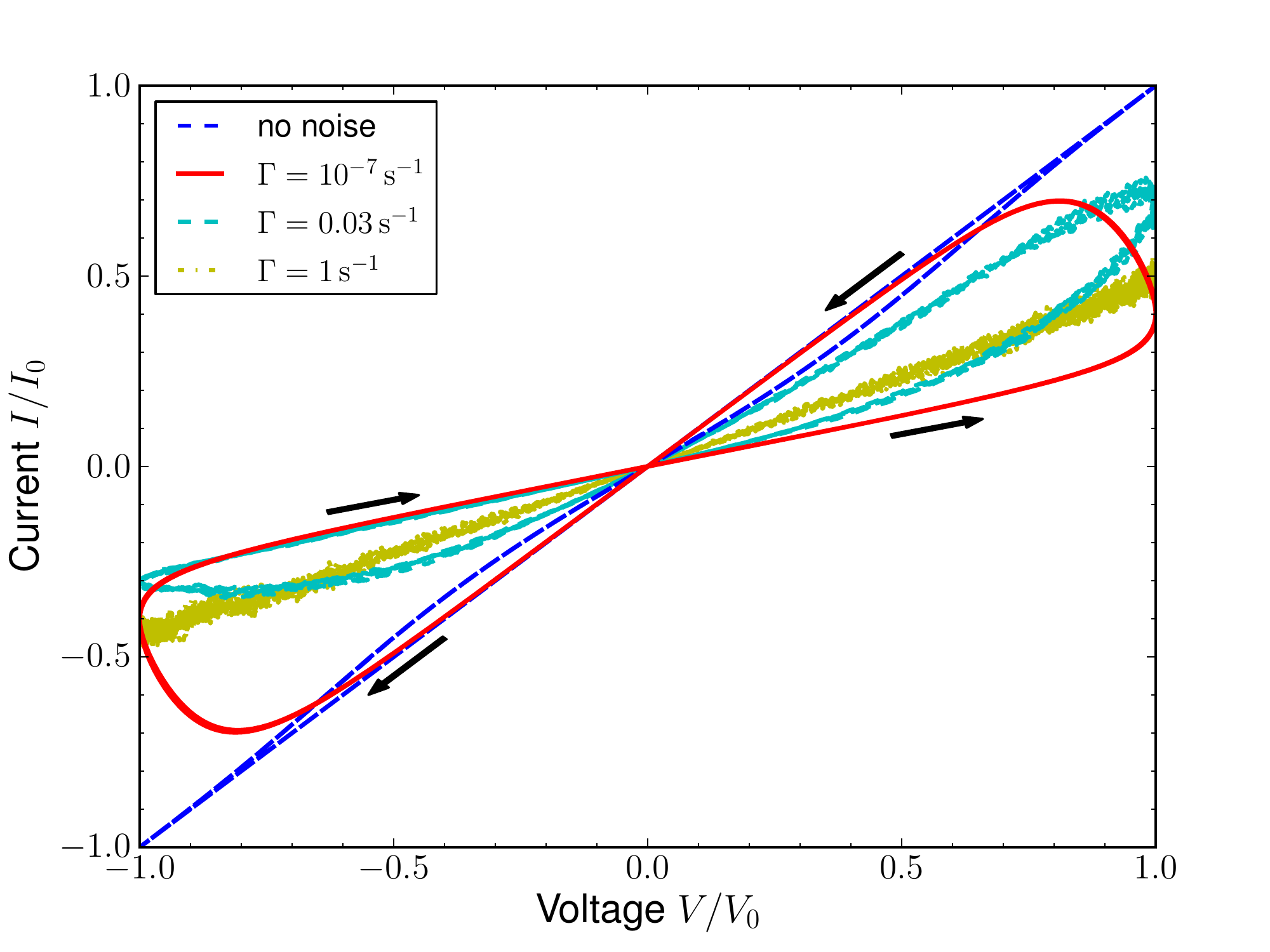}
 \caption{Current-voltage characteristic of the memristor represented in Fig.~\ref{fig:model} averaged over $100$
realizations of the stochastic process and for different noise strengths. Here $I_0 = V_0/R_{\tbox{on}}$ is the maximum possible current
through the system. The other parameters are the same as in Fig.~\ref{fig:memristance}.
}
\label{fig:IV}
\end{figure}

\begin{figure}[t]
 \centering
 \includegraphics[clip, width=0.95\hsize]{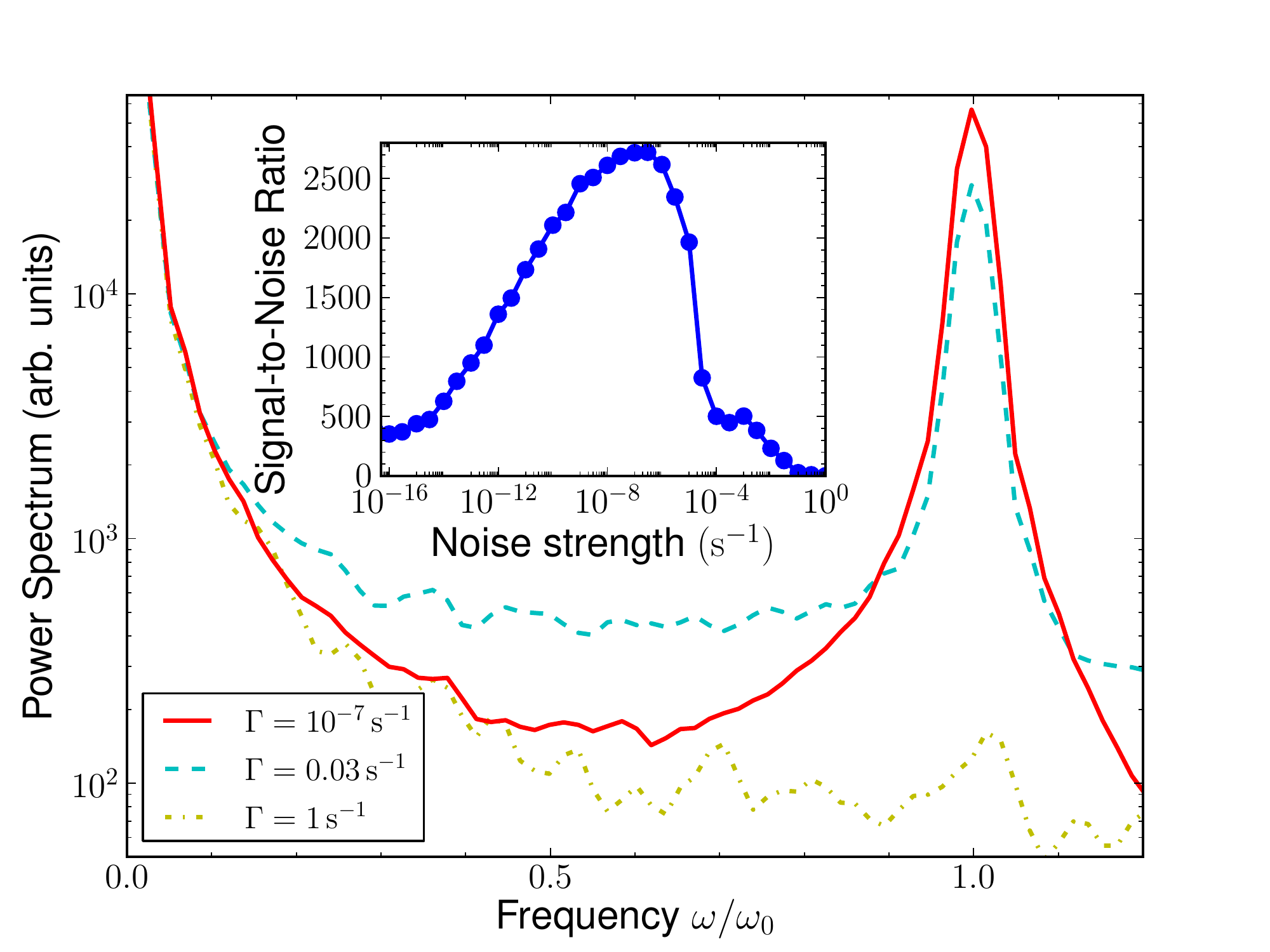}
 \caption{Power spectrum averaged over $100$
realizations. The height of the peak at $\omega = \omega_0$
is a measure of synchronization between the applied voltage and the
noise. Inset: The signal-to-noise ratio at $\omega = \omega_0$ is
plotted vs the noise strength and its maximum defines the stochastic
resonance.}
\label{fig:Spectrum}
\end{figure}

Without any noise the size of the hysteresis in the current-voltage
characteristic depends on the applied voltage frequency. In particular, at low frequencies ($\omega \ll 2\pi/t_0$) the
variation of memristance is very small (see Fig.~\ref{fig:memristance}) with a consequent small hysteresis loop (Fig.~\ref{fig:IV}).
The addition of a small amount of noise does not change this hysteresis loop considerably, but by increasing the noise strength we
find an optimal value of $\Gamma$ at which the hysteresis is considerably enhanced. In order to find this value we proceed
as follows. For every finite value
of the noise strength $\Gamma$ the stochastic differential equation
[Eq.~(\ref{eq:motion})] (along with the current and
Eq.~(\ref{eq:memristance})) is solved
(using the Euler-Maruyama method \cite{Kloeden:sde}) and the
boundary position $\{w(t)\}_\xi$ is
calculated for every realization of the stochastic process.
The power spectrum (defined as the squared absolute value of the
Fourier transform of $\{w(t)\}_\xi$) is then computed and averaged over
multiple realizations.
If the output $\{w(t)\}_\xi$ is well synchronized with the external periodic
voltage, the power spectrum exhibits a strong peak at the voltage frequency
$\omega_0$ (Fig.~\ref{fig:Spectrum}). The signal-to-noise ratio (SNR) is then defined
as the height of the peak
divided by the height of the noise background at frequency $\omega_0$. The SNR
attains its maximum for the noise strength at which the stochastic resonance
occurs (see the inset of Fig.~\ref{fig:Spectrum}). From our simulations we find that the maximum SNR is found for $\Gamma =
10^{-7}$s$^{-1}$ (Fig.~\ref{fig:Spectrum})~\footnote{Note that from Eqs.~(\ref{eq:diffgamma}) and
(\ref{eq:difftemp}) this value would correspond to a
temperature of $154\,\mathrm{K}$. However, the model we use here is too simplistic to give an accurate determination of
this number.}.
Fig.~\ref{fig:IV} indeed confirms that for this value of noise the hysteresis loop is the
widest, compared with the loops for other values of noise. In particular,
we find that stronger noise destroys the hysteresis loop as intuitively expected.

This picture can be explained qualitatively as follows. When the
frequency and amplitude of the external drive are low and the external noise introduced
into the system is zero or very weak, the boundary movement closely follows the
change in the applied voltage and sweeps a small region of the system~\footnote{Note that for other initial conditions wider loops
could be obtained for zero noise.}. If the
noise is very strong, the boundary moves erratically between the edges of
the memristor with almost no dependence on the applied voltage. In both cases
the hysteresis loop is narrow.
Wide loops are obtained whenever the boundary sweeps wide regions following the
changes in the voltage. For intermediate values of noise one can expect the
noise to ''help'' the transitions of the boundary from one edge to the other. Whenever
the boundary is moving from, say, the left edge toward the right one, the
noise can randomly kick it to the right edge and on the way back kick it back
to the left edge providing an almost ideal periodic movement (see Fig.~\ref{fig:memristance}).

Finally, we have also checked the effect of
introducing colored noise into the system. We have considered noise with an
exponentially decaying autocorrelation function
\begin{eqnarray}
  \langle \xi(t) \rangle = 0, \ \
  \langle \xi(t) \xi(t') \rangle = \Gamma
\exp\left(-\frac{|t-t'|}{\tau_c}\right)
\end{eqnarray}
where $\tau_c$ is the correlation time. In this case, we have indeed found (not reported here) enhancement of memory
effects even at high frequencies $\omega \gg t_0$ with $\tau_c \sim t_0$. This is not too surprising since, loosely speaking, the addition of this type of
noise is reminiscent of the addition of an external driving with frequencies
$\omega \lesssim 1/\tau_c$.

\smallskip

\section{Conclusions}
We conclude by noting that the above predictions can be tested experimentally by using memory resistors such as those considered in Ref. \cite{Strukov:memristor} or any other memory element~\cite{pershin:memoryreview}.
In a real experiment the frequency and the noise
can be easily changed, the latter via, e.g., the external temperature. If we focus again on the model considered here, according to Eq.~(\ref{eq:difftemp}), the diffusion
exponentially depends on the temperature. Therefore, if vacancy diffusion is the main mechanism of memory in $\mathrm{TiO_2}$ thin films, the opening of the hysteresis should be strongly dependent
on the external temperature, at least for small enough frequencies. This can be checked by lowering the frequency of the applied voltage below
the optimal one for a wide hysteresis loop, tuning the temperature and observing the change in the shape
of the hysteresis. Since the mechanism of memory in this particular memristive system is still
under debate, with redox effects at the metal-$\mathrm{TiO_2}$ interface the
other most probable cause \cite{wu:redox, Jeong:redox}, the tunability of the
noise strength  provides a possible diagnostic
tool to distinguish between the different mechanisms at play.
In general, since in a
real system noise is always present, it is highly possible that the hysteresis curves found experimentally in several memory elements~\cite{pershin:memoryreview}
could result from the cooperative effect of the optimal frequency in the presence of noise.

%%%%%%%%%%%%%%%%%%%%%%%%%%%%%%%%%%%%%%%%%%%%%%%%%%%%%%%%%%%%%%%%

\acknowledgments

We thank M. Krems for useful discussions. We acknowledge support from the Department of Energy Grant No. DE-FG02-05ER46204 and University of California Laboratories.

%%%%%%%%%%%%%%%%%%%%%%%%%%%%%%%%%%%%%%%%%%%%%%%%%%%%%%%%%%%%%%%%
\bibliography{sme}
%%%%%%%%%%%%%%%%%%%%%%%%%%%%%%%%%%%%%%%%%%%%%%%%%%%%%%%%%%%%%%%%

%%%%%%%%%%%%%%%%%%%%%%%%%%%%%%%%%%%%%%%%%%%%%%%%%%%%%%%%%%%%%%%%

\end{document}